\newif\ifarxiv
\arxivtrue 

\newcommand{\defeq}{:=}
\ifarxiv
	\documentclass[12pt]{article}
	\usepackage[margin=1in]{geometry}
	\usepackage[slantedGreek]{mathpazo}
	\usepackage[pdftex]{graphicx}
	\usepackage{amsfonts}
	\usepackage{amssymb}
	\usepackage{amsthm}
	\usepackage{graphicx,float}
	\usepackage{amsmath}%
	\usepackage{hyperref}
	\usepackage{natbib}
\else
\fi
\usepackage{color,soul,marginnote}

\ifarxiv
\title{Generative Quantile Bayesian Prediction}
\author{	
	\makebox[.4\linewidth]{Maria Nareklishvili}\\
	\textit{\small  Department of Economics}\\
	\textit{\small  Stanford University}\\
	\and
	\makebox[.4\linewidth]{Nicholas Polson\footnote{Email: ngp@chicagobooth.edu}}\\
	\textit{\small  Booth School of Business}\\
	\textit{\small  University of Chicago}\\
	\and 
	\makebox[.4\linewidth]{Vadim Sokolov}\\
	\textit{\small  Department of Systems Engineering }\\
	\textit{\small  and Operations Research}\\
	\textit{\small  George Mason University}\\
}

\date{First Draft: June 7, 2025\\This Draft: \today}
\else
\fi

\graphicspath{{./fig/}}

\begin{document}
\ifarxiv
\maketitle
\begin{abstract}
\noindent   Prediction is a central task of machine learning. Our goal is to solve large scale prediction problems using Generative Bayesian Prediction (GBP).
By directly learning predictive quantiles rather than densities we achieve a number of theoretical and practical advantages. 
We contrast our approach with state-of-the-art methods including conformal prediction, fiducial prediction and marginal likelihood.  Our distinguishing feature of our method is the use of generative methods for predictive quantile maps.  We illustrate our methodology for normal-normal learning and causal inference. Finally, we conclude with directions for future research. 
\end{abstract}
\else
\fi

\vspace{0.1in}


\newpage
\section{Introduction}
Prediction is a key task for modern day machine learning. Our goal is to provide a large scale predictive framework.  \cite{efron2020prediction} describes how this is a challenge for modern statistical methods and how pure black-box \citep{breiman2001statistical} methods such as deep learning, neural networks and random forests can outperform traditional statistical regression methods. \cite{hill1968posterior} proposed  Bayesian inference approach that does not require specification of a prior, later \cite{lei2014distributionfree} extended it to the case of regression.  We focus on generative Bayesian methods that directly model the predictive quantile function and hence are  likelihood-\emph{and}-prior-free.
An important property in high-dimensional problems as was recently pointed out by \cite{ritov2025thomas}.  Our approach builds on the quantile inference framework developed by Parzen (2004, 2009) which we describe in detail. 

Our approach provides an alternative to conformal prediction methods used extensively in machine learning.  The basic insight of conditional generative modeling is that is can be performed by non-parametric quantile regression \cite{white1992nonparametric}. Specifically, we wish to find a family of quantile functions to describe the predictive distribution of an output variable $Y$ given $X$  given by
$$
F_{Y|X}^{-1} ( \tau | x ) : ( \tau , x ) \rightarrow \inf \{ t \in \mathbb{R}^d : F_{Y|X}( t | x ) \geq \tau \} 
$$
Our work also builds on the implicit quantile NN literature \cite{dabney2018implicit} and generative Bayesian modeling framework of \cite{polson2023generative}.

We wish to directly find the prediction rule \cite{breiman2001statistical}. We will circumvent the use of parameters. There are three ways of writing a conditional predictive expectation: (1) using densities, (2) using survival functions, and (3) using quantiles. 
\begin{align*}
	\hat Y(x) = & E(Y\mid X) = \int_{-\infty}^{\infty}yf(y\mid x)dy\\
	= & \int_{0}^{\infty} (1 - F_{Y\mid X}(x))dx \text{ if } Y>0\\
	= & \int_{0}^{1}F^{-1}_{Y\mid X}(x)dx
\end{align*}

In Bayes framework, the predictive density calculation requires evaluating the marginal density via integration
\[
	f_{Y\mid X}(x) = \int f_{Y\mid \theta}(y\mid \theta)(\theta\mid x)d\theta.
\]
It can be done via MCMC. In Generative AI \cite{polson2023generative} we directly find the quantile function map $F^{-1}_{Y|X}(\tau)$ as a neural network. 

While our approach is prior-free and likelihood-free in the sense that we do not explicitly specify these quantities, it is important to recognize that implicit assumptions are embedded in the methodology. The choice of neural network architecture, loss function, and regularization implicitly defines a class of conditional distributions we can represent. Understanding the properties of this implicit class remains an important area for theoretical investigation. The connection to Wang's distortion functions, illustrated in our normal-normal learning example, provides valuable theoretical insight. This shows that our quantile updates can be interpreted as applying a distortion function to transform prior distributions into posterior distributions. Characterizing the class of distortion functions that can be learned by neural networks would provide theoretical guidance for architecture design.

Consider a pure prediction problem. The adjective ``pure" is justified by an algorithm's focus on prediction and the neglect of estimation and attribution \cite{efron2020prediction}. The basic strategy is simple: to go directly for high predictive accuracy and not worry about surface plus noise models. This has some striking advantages
and some drawbacks, too. Specifically, suppose that we have input-output pairs $ D = ( X_i , Y_i ) $. Pure prediction algorithms can be very different from each other. The least intricate and easiest to describe is random forests \cite{breiman2001statistical}. Recent research has focused on conformal prediction and prediction-powered inference. 

The basic \textbf{prediction problem} can be described as follows. One has training data $ Z_i = ( Y_i , X_i )_{i=1}^n $. The variables are assumed to be generated in a  stochastic fashion and the joint distribution assumed to be exchangeable.   The future observations consist of unobserved outcomes $ Z_i = ( \cdot , X_i )_{i=n}^N $ and observed covariance. We are interested in the case where $ N \geq n $.  The question is how do you predict the missing $ Y$'s?

The goal is to find a distribution-free prediction rule together with uncertainty quantification.  We model  i.i.d data $ Z_1 , \ldots , Z_n \sim P $ where $ Z_i =(X_i , Y_i ) \in \mathbb{R}^d \times \mathbb{R} $ comprises a response and a $d$-dim vector of features (aka. predictors).
Let  the regression surface be
$$
\mu(x ) = E( Y| X=x ) , x \in \mathbb{R}^d
$$
The goal is to  predict $Y_{new} $  at a new set of feature $ x_{new} $ with no assumptions on $ \mu , P $!

There have been a number of approaches to this problem. Recently, \cite{zrnic2024crosspredictionpowered} propose a prediction-powered inference framework. This is based on conformal prediction intervals which take the form
$$
\mathbb{P} \left ( Y_{new} \in C( x_{new} ) \right ) \geq 1 -\alpha 
$$
In turn this is very similar to \cite{cox1975prediction} fiducial prediction interval idea. \cite{barndorff-nielsen1996prediction} provide asymptotic for such methods and a general discussion. 

In classification, we have an  output  rule $ f(x, d)$  that, for any predictor vector $x$, yields a
prediction rule 
$ \hat{y} = f(x, d) $. 
The hope is that the apparent error rate of the rule, for classification problems the proportion of cases where $err  = \# ( \hat{y}_i \neq y )   / n $ is small. More crucially, we hope that the true error rate is small, 
$$
Err = E  ( f(X, d) \neq Y )  \; \; {\rm where}  \; \; ( X,Y) \sim P 
$$
Now a central tenet of Bayesian inference is coherence: the requirement that all probabilistic statements and predictions should be internally consistent and derived from a single joint probability model. This coherence is achieved by specifying a prior distribution over the parameters, which is then updated via Bayes' theorem in light of observed data.  We build on the statistical framework of \cite{parzen2004quantile,parzen2009last} who shows that  Bayes rule for quantiles is composite (superposition) of functions.
As such we can replace the prediction problem by one of non-parametric distribution matching \cite{lu2025likelihoodfree} using quantile neural networks \cite{polson2024generative,polson2023generative}.

\paragraph{Quantile Bayesian Predictive} This has been widely used in distributional decision-making \cite{dabney2017distributional,polson2024generative} and econometrics \cite{chernozhukov2010quantile,chernozhukov2021distributional}. Although classical quantile methods were developed for univariate variable \cite{koenker2005quantile}, recently there were several methods proposed for multivariate case \cite{carlier2016vector,kim2025deep}. They require the use of neural networks to train a transport map from estimated. 

The posterior distribution for parameters $\pi_F(\theta | D_{obs})$ (where $D_{obs}$ is the observed data) can be used to form a predictive distribution, denoted by  $p(y_{new} | x_{new} , D_{obs})   $, for a new observation $Y_{new}$ given $ x_{new} $: Let $ \theta$ be an unobserved latent variable (a.k.a. parameter) 
$$
F_{Y_{new} | x_{new} , D_{obs} } ( y )   = E_{ \theta  | x_{new} , D_{obs} } (  F_{ Y_{new} | \theta  } ( y ) )  = 
    \int_\Theta  F_{ Y_{new} | \theta  } ( y )  d F_{ \theta | x_{new} , D_{obs} } ( \theta ),
$$
where $F_{ Z_{new} | Z } $ is the conditional predictive distribution.

\paragraph{Bayesian Predictive Density} The posterior distribution for parameters $\pi_F(\theta | D_{obs})$ (where $D_{obs}$ is the observed data) can be used to form a predictive distribution, denoted by  $p(y_{new} | x_{new} , D_{obs})   $, for a new observation $Y_{new}$ given $ x_{new} $:
\[
p_F (y_{new} | x_{new} , D_{obs}) = \int p(y_{new} | \theta, x_{new} ) \pi_F(\theta | D_{obs}) d\theta
\]
where $p(y_{new} | \theta)$ is the model's likelihood for a new observation given the parameter $\theta$. From this predictive distribution, one can construct prediction intervals, for instance, by taking the $(\alpha/2)$ and $(1-\alpha/2)$ quantiles.

From computational point of view, Bayesian inference can be expensive as it requires computing high-dimensional integral numerically. We propose an alternative approach that leads to coherent inference and allows quantifying uncertainty. We use quantile neural networks. Those are composite maps that allow to avoid integration required for Bayesian inference. By bypassing the density functions and working directly with quantiles allows us to develop computationally efficient procedures to perform probabilistic inference.

\paragraph{Conformal prediction} This is a statistical technique that provides a flexible framework for constructing prediction intervals or sets with a specified level of confidence, regardless of the underlying data distribution. It operates by assessing the conformity of new data points with a set of previously observed data, using a nonconformity measure to quantify how unusual or typical a new observation is relative to the existing data. By leveraging the concept of exchangeability, conformal prediction ensures that the constructed prediction intervals or sets have valid coverage probabilities, meaning they contain the true value of the response variable with a pre-specified probability, even in finite samples. This makes conformal prediction a powerful tool for uncertainty quantification in machine learning and statistical inference, as it provides distribution-free, model-agnostic guarantees on the reliability of predictions.
\cite{angelopoulos2022private} and \cite{angelopoulos2023predictionpowered} describes the use of conformal predictive methods.  We show that generative  quantile Bayesian methods are a natural approach for predictive inference. \cite{polson2023generative} developed these methods for inference, \cite{polson2024generative} for reinforcement learning.
However, while conformal maps provide uncertainty quantification they do not necessarily lead to coherent inference and might imply a model with a non-existing prior.

\paragraph{Prediction vs Fiducial} 

Prediction intervals derived from fiducial predictive distributions do not automatically possess the strong, finite-sample, distribution-free marginal coverage guarantees that characterize conformal prediction intervals. The frequentest coverage of fiducial prediction intervals can be complex and may depend on the specific model and the properties of the fiducial argument used \citep{liu2024inferential}. In contrast, conformal prediction achieves its guarantees by construction, through the calibration step using nonconformity scores on exchangeable data.

Quantile regression  models the conditional mean of a response variable given certain predictor variables, quantile regression models the conditional quantiles (or percentiles) of the response variable. This allows for a more comprehensive understanding of the relationship between variables, particularly when the conditional distribution of the response variable is non-normal or when there's interest in different parts of the distribution (e.g., the 10th percentile, median, 90th percentile). It is particularly useful for data with heterogeneous variance (heteroscedasticity) because it can model how the spread of the distribution changes with the predictors.
Generalized fiducial inference \citep{hannig2016generalized}. Posterior predictive checks \cite{box1980sampling} and \cite{sinharay2003posterior}. 
\paragraph{Bayes and Fiducial} Suppose that $ \exists \phi = u( \theta )  $ such that re-parameterisation
$$
\frac{ f ( x , z | \phi ) }{ \int  f ( x , z | \phi ) d \phi } = g ( \phi - h( x , z ) ) 
$$
Then the plug-in predictive is
$$
\hat{f} ( z | x ) = c( x ) \sup_{\theta \in \Theta }  \; f( x | \theta ) g ( z | \theta ) 
$$
Hence, as $ \sup_{\theta  }  \; f( x,z | \theta ) = \sup_{\phi}  \; f( x,z | \phi ) = \int  f ( x , z | \phi ) d \phi  \times  \sup_\phi g ( \phi - h(x,z ) )  $, we have 
$$
\sup_{\theta \in \Theta }  \; f( x,z | \theta ) = c \int f( x , z | \theta ) \left| \frac{du}{d \theta} \right| d \theta 
$$
\cite{hannig2016generalized} extends this to generative methods such as Deep Fiducial Prediction. Even though the method is likelihood and prior free there is of course an implicit prior. Specifically,  Hannig  shows that the prior is "empirical" Jeffreys. Observed vs expected information. Sandwich estimator.  Thus providing a nice interpretation of such default Bayes procedures. 

Essentially $ | du / d \theta | $ is the implicit prior. Empirical Jeffreys arises from Laplace approximation. Same idea for predictive. 

An alternative approach due to \cite{leonard1976alternative} known as reverse predictive Bayes uses a backwards application of Bayes theorem. 
With future data $z$ and current $x$,  a backwards application of Bayes theorem yields
$$
g(z | x) = \frac{ g(z | \theta ) \pi( \theta | x )  }{ \pi ( \theta | x, z) } \; \; \forall \theta 
$$
where $g(z | x)$ denotes the predictive distribution of $z$ given $x$.

The full posterior  $ \pi( \theta | z , x ) $ has a normal approximation as $N$ large with var-covariance matrix $ R(z,x) $.

This provides asymptotic approximations $ O( N^{-1} ) $ for predictive densities. See \cite{leonard1976alternative} and \cite{barndorff-nielsen1996prediction}.

Reid provides approximations for predictive quantiles!

\subsection{Estimation Methods}

\paragraph{Kernel methods} Bartlett  \cite{nadaraya1964estimating} and \cite{watson1964smooth} proposed the use  of kernels to estimate the regression function. The idea is to estimate the regression function $f(x)$ at point $x$ by averaging the values of the response variable $y_i$ at points $x_i$ that are close to $x$. The kernel is used to define the weights.\\
\\
The regression function is estimated as follows
\[
	\hat{f}(x) = \sum_{i=1}^n  y_i K(x,x_i)/ \sum_{i=1}^n K(x,x_i) ,
\]
where the kernel weights are normalized.\\
\\
Both Nadaraya and Watson considered the symmetric kernel $K(x,x') = K(\|x'-x\|_2)$, where $||\cdot||_2$ is the Euclidean norm. The most popular kernel of that sort is the Gaussian kernel:
\[
	K(x,x') = \exp\left( -\dfrac{\|x-x'\|_2^2}{2\sigma^2}\right).
\]
Alternatively, the 2-norm can be replaced by the inner-product:
$K(x,x')  =  \exp\left( x^Tx'/2\sigma^2\right) $.\\
\\
Kernel methods are supported by numerous generalization bounds which often take the form of inequalities that describe the performance limits of kernel-based estimators. A particularly important example is the Bayes risk for $k$-nearest neighbors ($k$-NN), which can be expressed in a kernel framework as:
$$
 \hat{f} ( x) =  \sum_{i=1}^N w_i y_i        \; {\rm where} \; w_i \defeq K( x_i , x ) /  \sum_{i=1}^N K( x_i ,x )   
$$ 
\cite{schmidt-hieber2024generative} shows how to use kernel methods for generative methods. 

\subsection{Quantile Neural Networks}
Quantile neural networks extend traditional neural networks to estimate conditional quantiles rather than just conditional means, making them particularly valuable for heteroscedastic data where variance changes across the feature space. This approach combines the flexibility of neural network architectures with the statistical properties of quantile regression \citep{koenker2005quantile}.

The core of quantile neural network regression is the pinball loss (also called quantile loss or check function). For a given quantile level $\tau \in (0,1)$, the loss for a prediction $\hat{y}$ and actual value $y$ is:
$\rho_\tau(u) = u(\tau - \mathbb{I}_{u < 0}) $
where $u = y - \hat{y}$ is the residual and $\mathbb{I}_{u < 0}$ is the indicator function. This can be written explicitly as:
\[
L_{\tau}(y, \hat{y}) = \begin{cases}
\tau \cdot (y - \hat{y}), & \text{if } y \geq \hat{y} \\
(1 - \tau) \cdot (\hat{y} - y), & \text{if } y < \hat{y}
\end{cases}
\]
The pinball loss is convex and provides asymmetric penalties that encourage the network to learn the $\tau$-th conditional quantile. When $\tau = 0.5$, this reduces to the mean absolute error, corresponding to median regression.

We use conditional quantile neural networks (a.k.a. implicit quantile neural networks IQN) \cite{dabney2018implicit}. We wish to estimate a function
$$
Q_\tau ( x) = \inf \left \{ y \in \mathbb{R} : F_{Y|X} ( y |x ) \geq \tau \right \}, 
$$
where $ F_{Y|X} ( \cdot | \cdot ) $ is the conditional cdf of $ Y | X=x $. We estimate $ Q_\tau $ by minimizing the empirical quantile loss over a model class $ \mathcal{Q} $, namely
$$
\hat{Q}_\tau = \arg \min_{ Q \in \mathcal{Q}  } \; \sum_{i=1}^N \rho_\tau ( y_i - Q ( x_i ) ) 
$$
where $ \rho_\tau (u) = \max ( \tau u , ( \tau -1 ) u ) $ is the check loss.

A quantile neural network  architecture typically consists of:
\begin{enumerate}
    \item \textbf{Input layer}: Takes the feature vector $x \in \mathbb{R}^d$
    \item \textbf{Hidden layers}: Multiple densely connected layers with non-linear activation functions (typically ReLU or $\tanh$)
    \item \textbf{Output layer}: Either produces a single output for one specific quantile, or multiple outputs for different quantile levels $\tau_1, \tau_2, \ldots, \tau_K$
\end{enumerate}

For multiple quantile estimation, the optimization objective becomes:
\[
\min_{\theta} \frac{1}{N} \sum_{i=1, k=1}^{N, K}  L_{\tau_k}(y_i, f_{\theta, \tau_k}(x_i)),
\]
where $f_{\theta, \tau_k}(x_i)$ is the network's prediction for the $\tau_k$-th quantile given input $x_i$, and $\theta$ represents all network parameters.

A critical constraint during training is ensuring quantile monotonicity:
\[
\hat{q}_{\tau_1}(x) \leq \hat{q}_{\tau_2}(x) \leq \ldots \leq \hat{q}_{\tau_K}(x) \quad \text{for} \quad \tau_1 < \tau_2 < \ldots < \tau_K
\]
This can be enforced through penalty methods \citep{cannon2011quantile} or monotonic network architectures \citep{weissteiner2021monotonevalue}.
Add Huber norm. Silver etc. 

\paragraph{QNN Prediction:}  Quantile neural networks integrate naturally with conformal prediction methods. The conformalized quantile regression procedure involves the following:

\begin{enumerate}
    \item Train a quantile neural network to predict $\hat{q}_{\alpha/2}(x)$ and $\hat{q}_{1-\alpha/2}(x)$
    \item Compute nonconformity scores on calibration data:
    \[
    R_i = \max\Big(\hat{q}_{\alpha/2}(x_i) - y_i,\, y_i - \hat{q}_{1-\alpha/2}(x_i)\Big)
    \]
    \item Find the empirical $(1-\alpha)(1+1/n)$-quantile $\hat{Q}$ of these scores
    \item Construct prediction intervals:
    \[
    C(x_{new}) = [\hat{q}_{\alpha/2}(x_{new}) - \hat{Q},\, \hat{q}_{1-\alpha/2}(x_{new}) + \hat{Q}]
    \]
\end{enumerate}
This approach, known as conformalized quantile regression (CQR) \citep{romano2019conformalized}, provides finite-sample coverage guarantees while maintaining the adaptive width properties of quantile regression.
A related class of methods are those based on fiducial prediction. 
While conformal prediction directly constructs prediction intervals $C(x)$ with guaranteed the frequentist coverage for future observations $Y_{new}$, fiducial inference primarily yields a distribution for parameters $\theta$. However, connections can be drawn:

The synergy between quantile regression and conformal prediction provides a powerful method for constructing prediction intervals that are both adaptive to local data characteristics (heteroscedasticity) and possess rigorous, distribution-free coverage guarantees.  \citep{romano2019conformalized}  conceptualized this as a two-stage process: first, use quantile regression to provide an initial, input-dependent \emph{map} to a prediction interval; second, use conformal prediction to calibrate this map to ensure the desired statistical coverage. This is distinct from "conformal mapping" in a geometric sense; here, "conformal" refers to the properties of the prediction methodology.
This yields an initial prediction interval for $x$:
\[
I_0(x) = [\hat{q}_{\tau_L}(x), \hat{q}_{\tau_U}(x)]
\]
The width of this interval, $\hat{q}_{\tau_U}(x) - \hat{q}_{\tau_L}(x)$, is inherently adaptive: it can vary with $x$, reflecting the model's estimate of local uncertainty or data spread. However, this initial interval $I_0(x)$ does not generally come with a finite-sample coverage guarantee for an arbitrary target level $1-\alpha$.

\section{Generative Bayesian Prediction}

For ease of notation,  we will consider the estimation of $ F_{Y|X} $. For our prediction problem $ Y= Z_{n+1} $ and $ X = ( X_1 , \ldots , X_n ) $. 
Given an exchangable sequence $Z_n = (Y_n,X_n)$ we would like to calculate
\[
p(Z_{n+1} \mid Z_1,\ldots,Z_n) = p( X_{n+1} | \hat{Z}_{n+1} ), 
\]
where $ \hat{Z}_{n+1} = E( Z_{n+1} | Z_1 , \ldots , Z_n ) $ is a predictive sufficient statistic

\paragraph{Parzen Conditional Quantile Approach} The key property of quantiles is the fundamental identity \cite{parzen2009last} which relates the conditional quantile function, $Q_{Y|X} ( \cdot ) $ to the marginal  one, $Q_Y ( \cdot ) $, namely  composite map 
$$
Q_{Y|X} ( \tau ) = Q_Y \left ( F_Y ( \tau ) | X = x \right ) .
$$
This follows from two fundamental identities: 
$$ Q_{ g(Y) } ( \tau ) = g ( Q_Y ( \tau ) ) \; \; {\rm and} \; \;   Y = Q_Y ( F_Y ( Y ) ) 
$$ 
with probability $1$ as $ Q_Y = F_Y^{-1} $. 
Apply this to the conditional random variable $ Y|X$.

\paragraph{Ranks} 
Define the ranks $ U_i = F ( Y_i | X_i = x_i ) $ that are i.i.d. $ U(0,1) $. So-called uniformization transformation.  Meanwhile, we can recover the data $Y_i $ by the map
$$
Y_i = F_{ Y_i | X_i }^{-1} ( U_i ) 
$$
Also have the relationship
$$
F_{Y | X}  ( u) = \int_0^1 \mathbb{I}_{ F^{-1}_{Y|X} ( \tau ) \leq u } d \tau 
$$
Order statistics have the following property, also true for quantiles,  if 
$$ 
U_i = F ( X_i , \phi )  \; \; {\rm then} \; \;  U_{(i)} = F ( X_{(i)} , \phi ) 
$$
Key is you can relate the conditional quantile function to marginal. Quantiles are deep learners.
Hence our estimation approach has an implicit prior.  

\paragraph{Discrete Case} In the discrete case \citep{parzen2009last}, transform  via, with $ p(y) = P(Y=y) $
$$
F_{mid} ( y ) = F(y ) - \frac{1}{2} p( y) 
$$
Then $ ( X_j , Y_j ) $ to ranks or pseudo-observations $ ( U_j , V_j ) $ via
$$
U_j = F_{ mid_X } ( X_j ) \; {\rm and} \; V_j = F_{ mid_Y } ( Y_j )
$$
Then use a density/quantile method  to calculate $ V|U $.

\paragraph{Cosine Embedding} Our approach based on directly learning the predictive quantile function $ F_{ Y | \hat{Y} }^{-1} $
using the cosine-embedding transformation.

\vspace{0.1in} 

Generative methods solves this  as follows. Let $ \tau \sim P_\tau $ be a base measure for a latent variable, $\tau$, typically a standard multivariate normal or  a multivariate vector of uniforms. 
The goal of generative methods is to characterize the predictive measure $ P_{Y|X} $ from the training data
$ ( X_i , Y_i )_{i=1}^N \sim P_{X,Y} $. 

A deep learner is used to estimate $ \hat{f} $ via the non-parametric regression
 $ Y = f(X, Z ) $ . The deep learner is estimated via a NN from the triples  $ ( X_i , Y_i , Z_i)_{i=1}^N \sim P_{X,Y} \times P_Z$.
of the input given the output.  The multivariate  non-parametric  regression $ Y = f(X, \epsilon) $  provides a method for estimating the conditional mean. Typically, estimators, $ \hat{f} $, include KNN and Kernel methods. Recently, deep learners have been proposed and the theoretical properties of superpositions of affine functions (a.k.a.
ridge functions) have been provided (see \cite{montanelli2020error}  and \cite{schmidt-hieber2020nonparametric}).

The following is essentially a version of de Finetti's theorem, see \cite{kallenberg1997foundations}.

\vspace{0.1in}

\paragraph{Predictive Noise Outsourcing Theorem} If $ (X , Y ) $ are random variables in a Borel space $  ( \mathcal{X} , \mathcal{Y} ) $ then there exists an r.v. $Z$ 
which is independent of $ Y$ and a function $ G^\star : [0,1] \times \mathcal{X} \rightarrow \mathcal{Y} $ 
$$
(X , Y )  \stackrel{a.s.}{=} (X ,  G^\star ( Z , X  ) )
$$
Hence the existence of $G^\star$  follows from the noise outsourcing theorem \cite{kallenberg1997foundations}. 

Furthermore, if there is a statistic $S(X)$ with 
$ X \bot Y | S(X) $, then
$$
Y | X    \stackrel{a.s.}{=}  G^\star ( Z , S(X)  ) )
$$
The role of $S(X)$ is to perform dimension reduction in $n$, the dimensionality of the signal.   $S(X) $ can be estimated optimally via a deep neural network as the conditional mean $ \hat{S} ( X) = E ( Y | X ) $ as shown by \cite{schmidt-hieber2020nonparametric}.

To fix notation, let $ Z = ( Z_1 , \ldots , Z_n ) $ be a vector of signals.  In the inductive problem, we have $  Z = ( Z_1 , \ldots , Z_n ) $ and $ Y= Z_{n+1} $ and we are simply characterising the conditional predictive
where $ Z_{n+1} = ( Y_{n+1} , X_{n+1} ) $ and
$$
p( Z_{n+1} | Z_1 , \ldots , Z_n ) 
$$
 We wish to find the conditional quantile function $ F_{ Y_{n+1} | X_{n+1} = x_{n+1} }$ via its quantile function. 

\paragraph{Architecture Design} 

In the non-regression setting, we have the map 
\[
Y_{n+1} = H\left\{S(Y_1,\ldots,Y_n)\right\}.
\]
When the problem is high-dimensional and $n$ is large, it becomes hard to formulate a prior based on observed $Y_1,\ldots,Y_n$. Further, calculating a likelihood might not be possible. Thus, replacing both prior with map $S$ (minimal sufficient statistics) and the likelihood-prior product with map $H$, we solve the problem of prior specification. Further, even if we can formulate a prior and have a tractable likelihood, calculating the posterior via MCMC can be prohibitively expensive. Replacing the posterior calculations with direct map evaluation of $H$ and $S$ makes problem computationally tractable.

\paragraph{Predictive Bayes Sufficiency}

There are many predictive sufficient statistics including those for exponential families and dimension reduction methods \citep{ressel1985finettitype}. 
 
 Predictive sufficiency directly assesses the conditional predictive $ p_{n+1}  ( y_{n+1} | y_1 , \ldots , y_n ) $ without recourse to parameters or prior distributions.
Predictive exchangeability and sufficiency has characterized predictive rules. \cite{diaconis1983quantifying} and \cite{ressel1985finettitype} show how you can recover the prior from the predictive sufficient statistic.  This has led to Polya Urn priors \citep{muliere1998extending} and the literature on NP-Bayes based on Dirichlet process.
\cite{vovk2017nonparametric} proposed Dempster-Hill procedure  based on the work of Jeffreys who proposed a predictive rule called $A_{(n)}$. \cite{dempster1963direct} calls these direct probabilities rather than fiducial ones. One can make interval probabilistic statements about future observations.  \cite{lane1978diffuse} shows problems with $ A_{(n)} $ and hence conformal predictive rules using the notion of non-conglomerability.

 \paragraph{de Finetti:} We need to assess the predictive distribution  $ p( Y_{n+1} \leq y_{n+1}  | y_1 , \ldots , y_n ) $.  
A key concept will be that of  exchangeability and prediction sufficiency.  Exchangeability ensues that the joint distribution is invariant to order and predictive sufficiency allows for dimension reduction in the predictive distribution. Let  $ \hat{Y} _{n+1} ( y_1 , \ldots , y_n ) $ be a predictor of $Y_{n+1} $ such as the conditional mean.
The assumption of exchangeability (also used in conformal prediction) allows one to expresses the predictive density as a marginal over parameters. Exchangeability implies we can \emph{act as if} there exists a parameter $ \theta $ and we update with Bayes rule. Alternatively, we can directly use predictive sufficiency and avoid parameters.  Our approach is related to the latter where we simply estimate the predictive quantile function. 

The conditional likelihood, under $ m$, is given by $ f_\theta(y) =  \prod_{i=1}^n f_\theta ( y_i ) $. 
Given a prior measure, $ \Pi ( d F ) $,  over $ \mathcal{F} $ the set of distributions, we can calculate the predictive density
$$
p  ( Z_{n+1} | Z_1 , \ldots , Z_n ) = \int f (y) \Pi_n ( d F ) \; {\rm where} \; \Pi_n ( d f ) = \frac{ \prod_{i=1}^n f( y_i ) \Pi( d f ) }{  \int  \prod_{i=1}^n f( y_i ) \Pi( d f ) }
$$
Under the family, $ f_\theta $, we can calculate the parameter posterior as
$$
p( \theta | y ) =  \frac{ \prod_{i=1}^n f_\theta ( y_i ) p(\theta)  d \theta }{  m(y) } \;  {\rm where} \; m(y) = \int f_\theta (y) p( \theta ) d \theta 
$$
Here $ p( \theta ) $ is a prior distribution over parameters and $ m(y)  $ is the marginal distribution of the data implied by the model. 

\cite{fong2023martingale} give a predictive sampling perspective rather than focusing on prior distribution over parameters using the argument that uncertainty in parameters arises mainly from missing data (unobservable future observations)

\paragraph{Neural Network Estimation}

There are several methods for neural network estimation of conditional quantile functions. \cite{white1992nonparametric} provides theoretical foundations for nonparametric conditional quantile estimation and establishes consistency results in the econometrics literature. Building on this foundation, \cite{polson2023generative} develop approaches using ReLU networks, following the theoretical framework of \cite{schmidt-hieber2020nonparametric}. These methods also incorporate kernel-based techniques similar to those used in approximate Bayesian computation (ABC) with local windowing.

\cite{kim2025deep} propose deep learning methods for multivariate quantile regression that extend classical univariate approaches to handle complex multivariate dependencies. Their framework uses neural networks to learn conditional quantile functions directly, avoiding the need for explicit distributional assumptions.

The approach of \cite{lu2025likelihoodfree} focuses on non-parametric distribution matching, which shares similarities with ABC methods. The key insight is that the infinite-dimensional problem can be reduced by conditioning on indicator functions of the form $\mathbb{I}(D(\cdot, \cdot) < \epsilon)$, where $D$ represents a distance measure between predictive densities.

A crucial consideration is the choice of distance measure $D$ between predictive densities. The earth mover distance provides one natural metric for comparing distributions. When working with samples from distributions, the question becomes how to effectively measure distances, leading to the nonparametric density estimation problem. While \cite{bishop1994mixture} addresses this challenge for density estimation, working directly with quantiles offers computational and theoretical advantages.

\section{Applications}

\paragraph{Efron Example}  A motivating normal distribution with unknown expectation $ \theta $, 
$$
( y_1 , \ldots , y_n |  \theta ) \sim N( \theta , 1 ) 
$$
and consider estimating µ with either the sample mean  $ \bar{x} $ or the sample median $x_{0.5}$. As far
as squared error is concerned, the mean is an overwhelming winner, being more than half
again more efficient,
$$
 \frac{E ( ( x_{0.5} - \theta )^2  )  }{E ( ( \bar{x} - \theta )^2 ) } = 1.57
$$
Suppose instead that the task is to predict the value of a new, independent realization
$$
 \frac{E ( ( x_{0.5} - x )^2  )  }{E ( ( \bar{x} - x )^2 ) } = 1.02
$$
The mean still wins but only by $2$\% now. The reason, of course, is that most of the prediction error comes from the variability of $X$, namely $ \sigma^2 =1 $ which neither  can cure. This imagines that we have a single new observation to predict. Suppose instead that we have m new
observations. Suppose we need to predict the mean of the next $m$ observations, then this will become an estimation problem!

\paragraph{Normal Learning}

Consider the predictive quantile function for a mixture of normal distributions. Rather than working with density mixtures, we find it advantageous to work directly with quantile mixtures. This approach leverages the Wang distortion map framework for quantile updating.

For a two-component mixture of normal distributions, we can demonstrate how quantiles update through the learning process. Following \cite{shen2002prediction}, the predictive quantile function takes the form:
\begin{align*}
Q ( y^\star , y_n ) & = \int_{ - \infty}^\infty \Phi \left ( \frac{ y^\star - \theta }{\sigma} \right ) d \Phi \left ( \frac{ \theta - \bar{y}  }{\sigma/\sqrt{n} } \right )\\
& = \Phi \left ( \frac{ y^\star - \bar{y} }{\sigma \sqrt{ 1 + 1/n} } \right )
\end{align*}

This formulation connects naturally to the distortion deep learning framework of \cite{wang2000class}, which provides a neural network approach for quantile updating in Bayesian prediction problems.

Under non-informative prior, $ y^\star = \theta + \epsilon $ and $ \theta = \bar{y} + \nu $, we  can integrate out $ \theta $. Same cdf calculations.

For the purpose of illustration, we consider the normal-normal learning model.  We will develop the necessary quantile theory to show how to calculate posteriors and expected utility without resorting to densities.  Also, we show a relationship with Wang's risk distortion measure as the deep learning that needs to be learned. 

Specifically, we observe the data $ y = ( y_1,\ldots, y_n)$ from the following model
\begin{align*}
	(y_{n+1} \mid \bar{y} ) & \sim N(\bar{y}, \sigma^2 + \nu^2) \\
\end{align*}
Hence, the summary (sufficient) statistic $S(y) = \bar y$. A remarkable result shows that we can learn $S$ independent of $H$ simply via OLS. 

Given observed samples $y = (y_1,\ldots,y_n)$, the posterior is then  $\theta \mid y \sim N(\mu_*, \sigma_*^2)$ with 
$$
\mu_* = (\sigma^2 \mu + \alpha^2s) / t, \quad \sigma^2_* = \alpha^2 \sigma^2 / t,
$$
where  
$$t =  \sigma^2 + n\alpha^2 \; \; {\rm and} \; \; s(y) = \sum_{i=1}^{n}y_i
$$
The posterior and prior CDFs are then related via the Wang distortion function
$$
1-\Phi(\theta, \mu_*,\sigma_*) = g(1 - \Phi(\theta, \mu, \alpha^2)),
$$
where $\Phi$ is the normal distribution function. Here
\[
g(p) = \Phi\left(\lambda_1 \Phi^{-1}(p) + \lambda\right),
\]
where
$$
\lambda_1 = \dfrac{\alpha}{\sigma_*} \; \; {\rm and} \; \; 
\lambda = \alpha\lambda_1(s-n\mu)/t .
$$
The proof is relatively simple and is as follows 
\begin{align*}
	g(1 - \Phi(\theta, \mu, \alpha^2)) & = g(\Phi(-\theta, \mu, \alpha^2)) = g\left(\Phi\left(-\dfrac{\theta - \mu}{\alpha}\right)\right)\\
	& = \Phi\left(\lambda_1 \left(-\dfrac{\theta - \mu}{\alpha}\right) + \lambda\right) =  1 - \Phi\left(\dfrac{\theta - (\mu+ \alpha\lambda/\lambda_1)}{\alpha/\lambda_1}\right)
\end{align*}
Thus,
$$
\sigma_* = \alpha/\lambda_1, \quad \lambda_1 = \dfrac{\alpha}{\sigma_*}
$$
and 
$$
\mu_* = \mu+ \alpha\lambda/\lambda_1, \quad \lambda = \dfrac{\lambda_1(\mu_* - \mu)}{\alpha} = \alpha\lambda_1(s-n\mu)/t
$$
\subsubsection*{Numerical Example}
Consider the normal-normal model with Prior $\theta \sim N(0,5)$ and likelihood $y \sim N(3,10)$. We generate $n=100$ samples from the likelihood and calculate the posterior distribution.

\begin{figure}[H]
\centering
\begin{tabular}{ccc}
\includegraphics[width=0.33\linewidth]{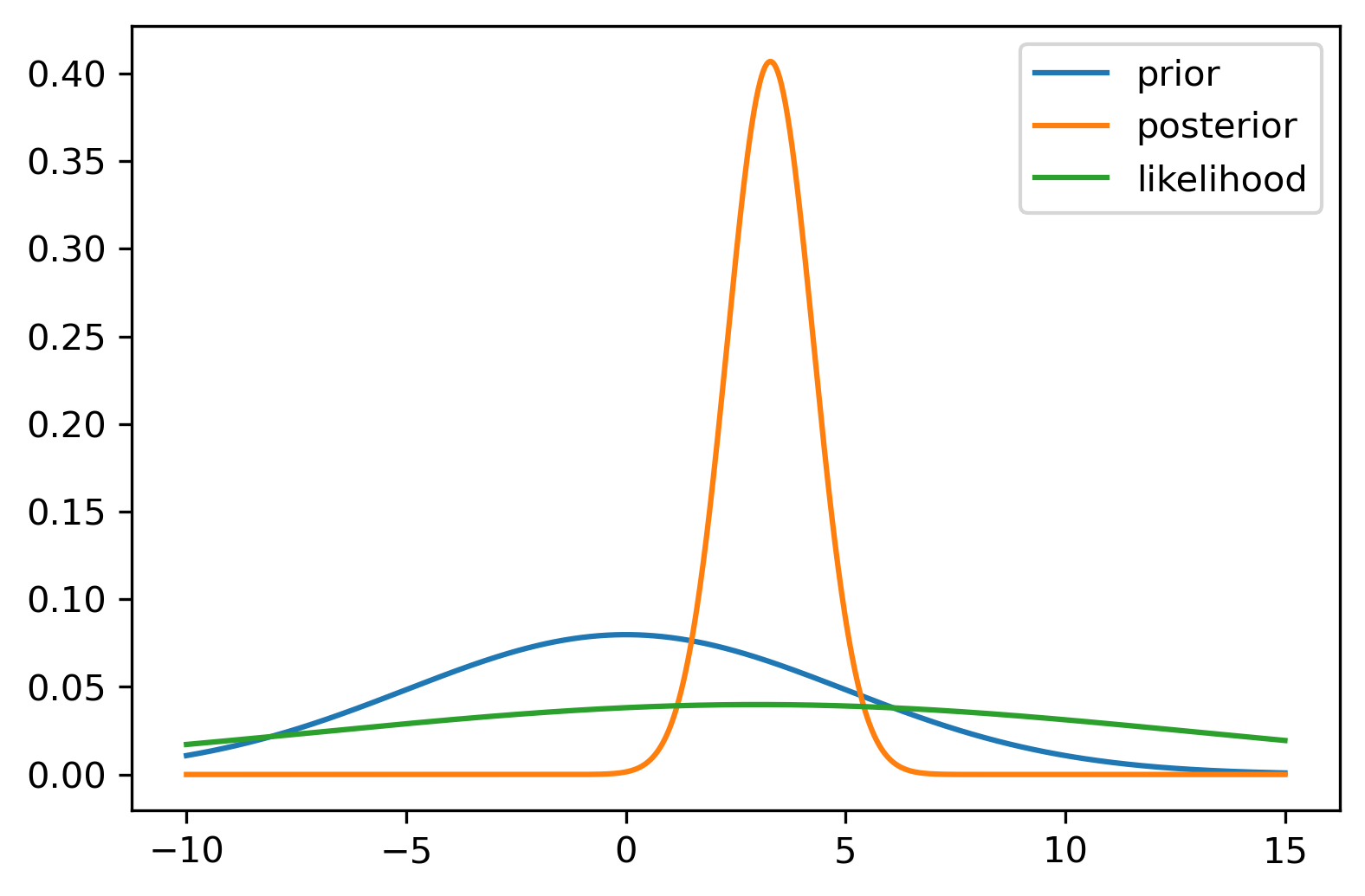} & \includegraphics[width=0.33\linewidth]{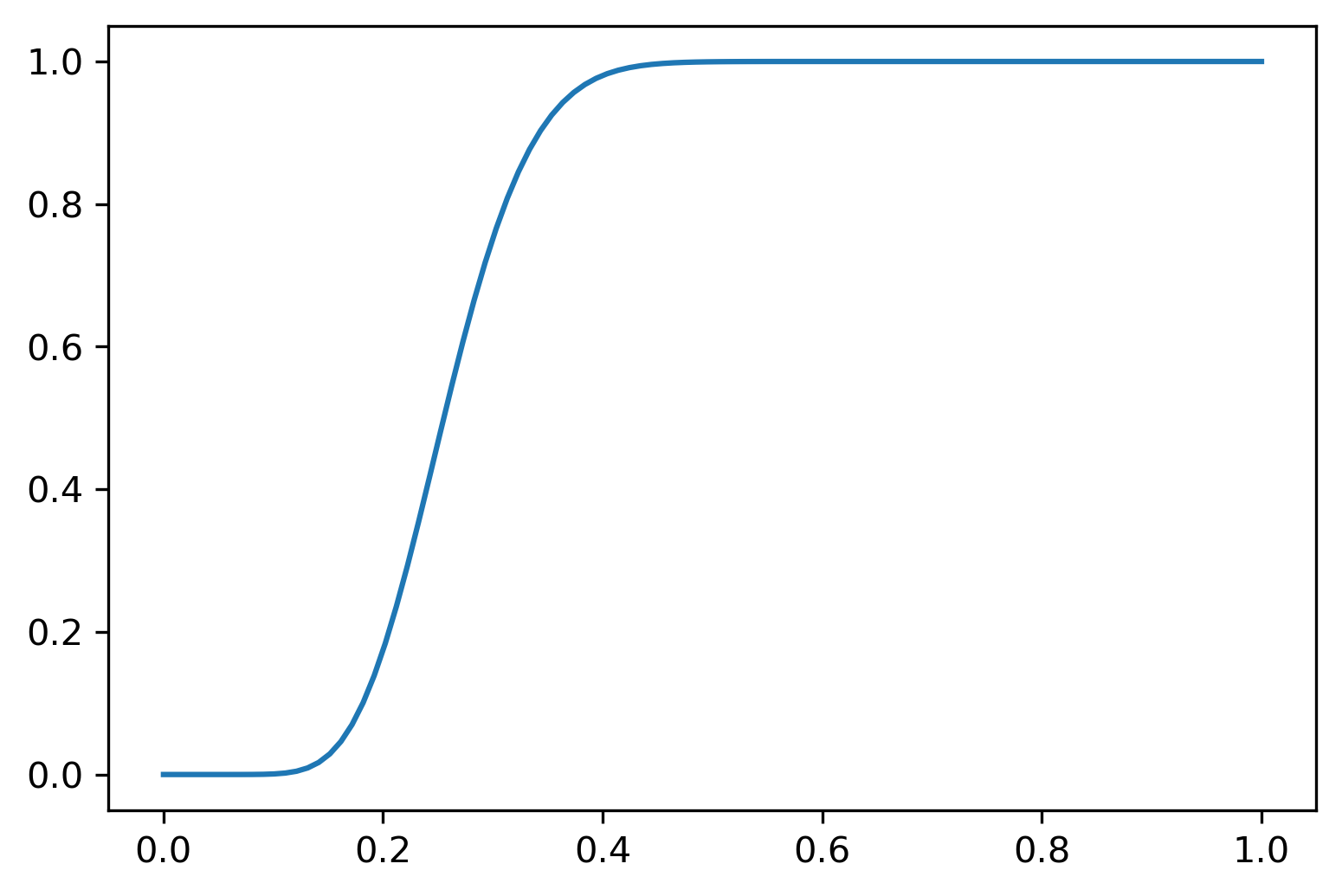} & \includegraphics[width=0.33\linewidth]{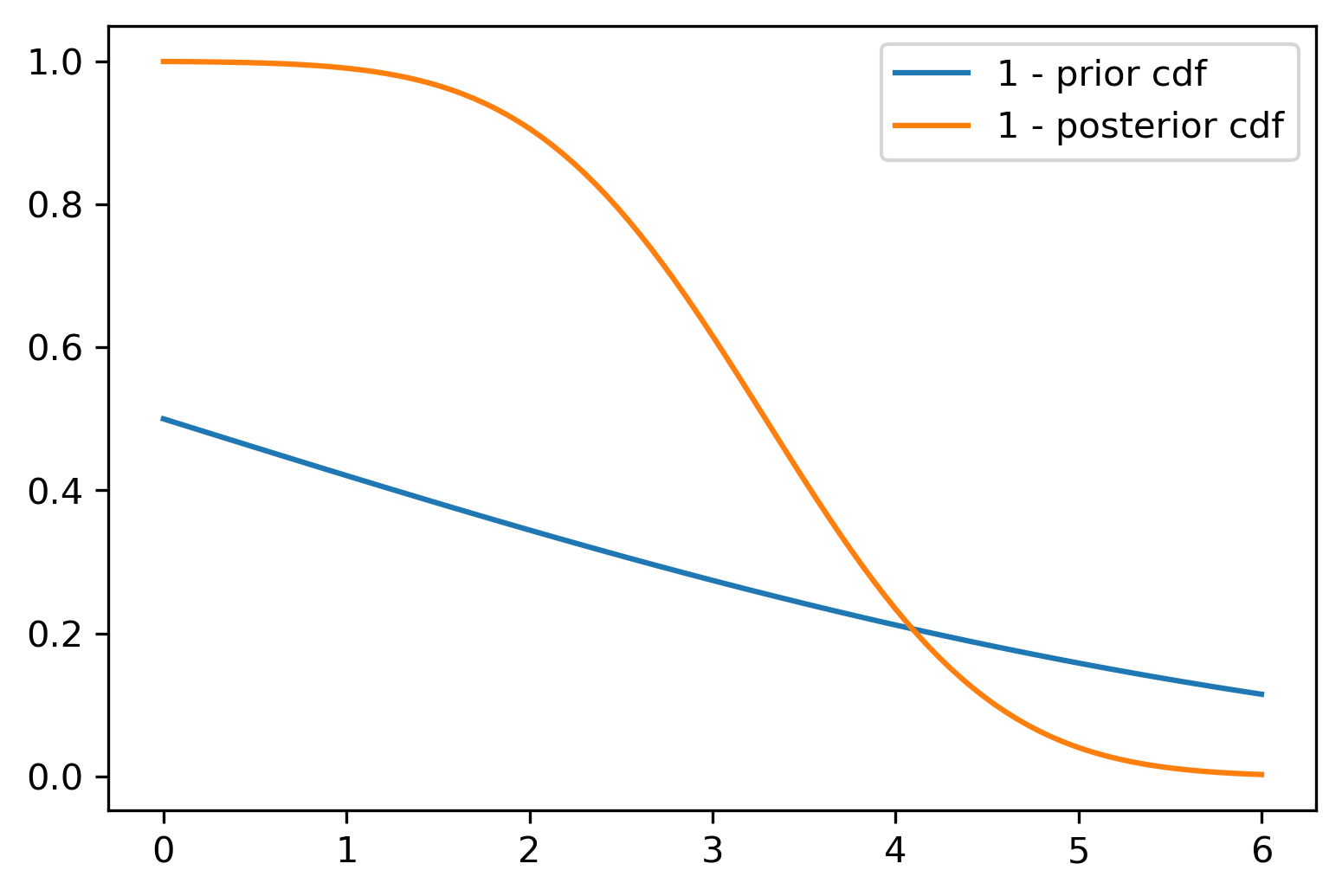}\\
(a) Model for simulated data & (b) Distortion Function $g$ & (c) 1 - $\Phi$\\
\end{tabular}
\caption{Density for prior, likelihood and posterior, distortion function and 1 - $\Phi$ for the prior and posterior of the normal-normal model.}
\label{fig:wang}
\end{figure}

The posterior distribution calculated from the sample is then $\theta \mid y \sim N(3.28, 0.98)$.

Figure \ref{fig:wang} shows the Wang distortion function for the normal-normal model. The left panel shows the model for the simulated data, while the middle panel shows the distortion function, the right panel shows the 1 - $\Phi$ for the prior and posterior of the normal-normal model.



\section{Discussion}

We have presented a generative Bayesian prediction framework that operates directly on predictive quantiles rather than densities, offering a novel approach to large-scale prediction problems. The fundamental insight underlying our methodology is Parzen's composite quantile identity, which shows that conditional quantiles update through function composition rather than integration. This mathematical property translates directly into computational advantages: neural networks excel at learning composite functions through their layered architecture, allowing us to circumvent the computationally expensive integration required for traditional Bayesian predictive densities while maintaining probabilistic coherence. Our approach provides a compelling alternative to conformal prediction methods while sharing their desirable property of being both likelihood-free and prior-free. However, unlike conformal prediction, which achieves distribution-free coverage through calibration while treating the prediction model as a black box, our generative quantile approach directly models the conditional predictive distribution and produces predictions consistent with an implicit generative model. This coherence property may be valuable in applications requiring probabilistic reasoning.

Compared to competing frameworks, our method offers distinct advantages. Traditional Bayesian predictive inference integrates over posterior distributions, requiring likelihood specification and prior elicitation, which faces significant challenges in high-dimensional settings. Fiducial inference, while also avoiding explicit priors, typically requires strong structural assumptions about the data generating process. Our quantile-based approach requires only the ability to learn the conditional quantile function from data through neural networks. From a computational perspective, once trained, prediction requires only a forward pass through the network—a constant-time operation that contrasts sharply with kernel methods or MCMC-based inference. The approach naturally handles heteroscedastic data and distributional asymmetries, scales to modern deep learning architectures, and adapts to local data characteristics by directly modeling conditional quantiles.

Despite its advantages, our approach faces several limitations. Like all neural network methods, quantile neural networks require careful hyperparameter tuning and may overfit in small-sample settings. Ensuring quantile monotonicity requires architectural constraints or penalties, and interpretation may be less transparent than for parametric models. Several promising directions for future research emerge: establishing finite-sample or asymptotic guarantees, extending to multivariate output spaces, developing connections to causal inference through quantile treatment effects, and incorporating domain knowledge through structured architectures. The fundamental insight that quantiles update through composition provides a solid foundation for future development in prediction with uncertainty quantification.

\bibliography{GenPrediction} 
\end{document}